\documentclass[11pt]{article}
\usepackage{DGfest}
\usepackage{graphicx}
\usepackage{times}

\newcommand{\be}{\begin{equation}}
\newcommand{\ee}{\end{equation}}
\newcommand{\bea}{\begin{eqnarray}}
\newcommand{\eea}{\end{eqnarray}} 

\newcommand{\de}{\partial}

\def\be{\begin{equation}}
\def\ee{\end{equation}}
\def\bea{\begin{eqnarray}}
\def\eea{\end{eqnarray}}

\begin{document}
\baselineskip 11.5pt
\title{COLOR SUPERCONDUCTING NEUTRAL MATTER}

\author{Roberto Casalbuoni}

\address{Department of  Theoretical Physics, University of Florence and\\
Sezione INFN of Florence}

%

\maketitle\abstracts{We describe the effects of the strange quark
mass and of the color and electric neutrality on the superconducing
phases of QCD.}

\section{Introduction}

It is now a well established fact that at zero temperature and
sufficiently high densities quark matter is a color superconductor
\cite{barrois,cs}. The study starting from first principles was done
in \cite{weak,PR-sp1,weak-cfl}. At baryon chemical potentials much
higher than the masses of the quarks $u$, $d$ and $s$, the favored
state is the so-called Color-Flavor-Locking (CFL) state, whereas at
lower values, when the strange quark decouples, the relevant phase
is called two-flavor color superconducting (2SC).

An interesting possibility is that in the interior of compact
stellar objects (CSO) some color superconducting phase may exist. In
fact the central densities for these stars could be up to $10^{15}$
g/cm$^{3}$, whereas the temperature is of the order of tens of keV.
However the usual assumptions leading to prove that for three
flavors the favored state is CFL should now be reviewed. Matter
inside a CSO should be electrically neutral and should not carry any
color. Also  conditions for $\beta$-equilibrium should be fulfilled.
As far as color is concerned, it is possible to impose a simpler
condition, that is color neutrality, since in \cite{Amore:2001uf} it
has been shown that there is no free energy cost in projecting color
singlet states out of color neutral states. Furthermore one has to
take into account that at the interesting density the mass of the
strange quark is a relevant parameter. All these effects, the mass
of the strange quark, $\beta$-equilibrium and color and electric
neutrality, imply that the radii of the Fermi spheres of the quarks
that would pair are not the same creating a problem with the usual
BCS pairing. Let us start from the first point. Suppose to have two
fermions of masses $m_1=M$ and $m_2=0$ at the same chemical
potential $\mu$. The corresponding Fermi momenta are
$p_{F_1}=\sqrt{\mu^2-M^2}$ and $p_{F_2}=\mu$. Therefore the radius
of the Fermi sphere of the massive fermion is smaller than the one
of the massless particle. If we assume $M\ll \mu$  the massive
particle has an effective chemical potential
$\mu_{\rm{eff}}=\sqrt{\mu^2-M^2}\approx \mu-{M^2}/{2\mu}$ and the
mismatch between the two Fermi spheres is given by
\be\delta\mu\approx\frac{M^2}{2\mu}\label{eq:3}\ee This shows that
the quantity $M^2/(2\mu)$ behaves as a chemical potential. Therefore
for $M\ll\mu$  the mass effects can be taken into account through
the introduction of the  mismatch between the chemical potentials of
the two fermions given by eq. (\ref{eq:3}). This is  the way that we
will follow in our study.

Now let us discuss  $\beta$-equilibrium. If electrons are present
(as generally required by electrical neutrality) chemical potentials
of quarks of different electric charge are different. In fact, when
at the equilibrium for $d\to ue\bar\nu$, we have \be
\mu_d-\mu_u=\mu_e\ee From this condition it follows that for a quark
of charge $Q_i$ the chemical potential $\mu_i$ is given by \be
\mu_i=\mu+Q_i\mu_Q\ee where $\mu_Q$ is the chemical potential
associated to the electric charge. Therefore \be\mu_e=-\mu_Q\ee
Notice also that $\mu_e$ is not a free parameter since it is
determined by the neutrality condition \be Q=-\frac{\de \Omega}{\de
\mu_e}=0\ee At the same time the chemical potentials associated to
the color generators $T_3$ and $T_8$ are determined by the color
neutrality conditions \be\frac{\de \Omega}{\de \mu_3}= \frac{\de
\Omega}{\de \mu_8}=0\label{eq:1}\ee

We see that in general there is a mismatch between the quarks that
should pair according to the BCS mechanism for $\delta\mu=0$.
Therefore the system might go to a normal phase since the mismatch,
as we shall see, tends to destroy the BCS pairing, or a different
phase might be formed. In the next Sections we will explore some of
these possible phases.

\section{Pairing Fermions with Different Fermi Momenta}

\begin{figure}[h]\begin{center}
  \includegraphics[height=.3\textheight]{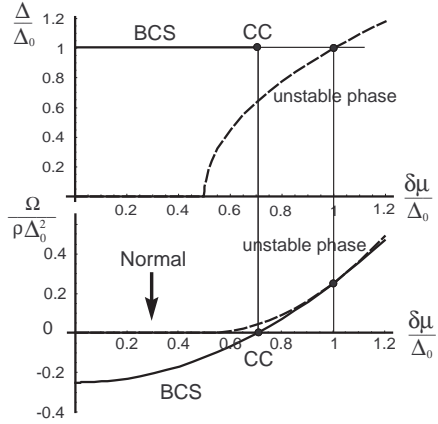}
  \caption{The two solutions of the gap equation with a mismatch
  $\delta\mu$. The continuous line is the BCS solution, the dashed
  one is called the Sarma solution.
  \label{fig:1}}\end{center}
\end{figure}

We start now our discussion considering a simple model with two
pairing quarks, $u$ and $d$,  with  chemical potentials
\be\mu_u=\mu+\delta\mu,~~~~\mu_d=\mu-\delta\mu\ee and no further
constraints. The gap equation for the LOFF phase at $T=0$ is given
by (see for example ref. \cite{Casalbuoni:2003wh}) \be
1=\frac{g}2\int\frac{d^3p}{(2\pi)^3}\frac{1}{\epsilon(\vec
p,\Delta)}\left(1-\theta(-\epsilon-\delta\mu)-\theta(-\epsilon+\delta\mu)\right)\label{gap}\ee
where \be \epsilon(\vec p,\Delta)=\sqrt{\xi^2+\Delta^2},~~~\xi=\vec
v_F\cdot(\vec p-\vec p_F)\ee with $\vec v_F$ and $\vec p_F$ the
Fermi velocity and Fermi momentum. The meaning of the two step
functions is that at zero temperature there is no pairing when
$\epsilon(\vec p,\Delta)<|\delta\mu|$. In other words the pairing
may happen only for excitations with positive energy. However, the
presence of negative energy states, as in this case, implies that
there must be gapless modes. When this happens, there are blocking
regions in the phase space, that is regions where the pairing cannot
occur.  The effect is to inhibit part of the Fermi surface to the
pairing giving rise a to a smaller condensate with respect to the
BCS case where all the surface is used. In the actual case the gap
equation at $T=0$ has two different solutions (see for instance ref.
\cite{Casalbuoni:2003wh}) corresponding to: a) $\Delta=\Delta_0$,
and b) $\Delta^2=2\delta\mu\Delta_0-\Delta_0^2$ where $\Delta_0$ is
the BCS solution of the gap equation for $\delta\mu=0$. The two
solutions are illustrated in Fig. \ref{fig:1}.

We see that the solution a) is always favored with respect to the
solution b) (called the Sarma phase \cite{sarma:1963sa}).
Furthermore the BCS phase goes to the normal phase at
\be\delta\mu_1=\frac{\Delta_0}{\sqrt{2}}\ee This point is called the
Chandrasekhar-Clogston (CC) point \cite{Chandrasekhar} (denoted by
CC in Fig. \ref{fig:1}). Ignoring for the moment that in this case,
after the CC point the system goes to the normal phase, we notice
that the gaps of the two solutions coincide at $\delta\mu=\Delta_0$.
This is a special point, since in presence of a mismatch the
spectrum of the quasi-particles is modified from\be
E_{\delta\mu=0}=\sqrt{(p-\mu)^2+\Delta^2}\to
E_{\delta\mu}=\left|\delta\mu\pm\sqrt{(p-\mu)^2+\Delta^2}\,\right|\ee
Therefore for $|\delta\mu|<\Delta$ we have gapped quasi-particles
with gaps $\Delta\pm\delta\mu$. However, for $|\delta\mu|=\Delta$ a
gapless mode appears and from this point on there are regions of the
phase space which do not contribute to the gap equation (blocking
region).

The gapless modes are characterized by \be E(p)=0\Rightarrow
p=\mu\pm\sqrt{\delta\mu^2-\Delta^2}\ee Since the energy cost for
pairing two fermions belonging to Fermi spheres with mismatch
$\delta\mu$ is $2\delta\mu$ and the energy gained in pairing is
$2\Delta$, we see that the fermions begin to unpair for $
2\delta\mu\ge 2\Delta$. These considerations will be relevant for
the study of the gapless phases when neutrality is required.

\section{The g2SC Phase}
The g2SC phase \cite{huang:2003ab} has the same condensate as the
2SC \be\langle
0|\psi_{aL}^{\alpha}\psi_{bL}^\beta|0\rangle=\Delta\epsilon^{\alpha\beta
3}_{ab 3},~~~\alpha,\beta\in SU_c(3),~~~a,b\in SU(2)_L\ee and,
technically, it is distinguished by the 2SC one by the presence of
gapless modes starting at $\delta\mu=\Delta$. In this case only two
massless flavors are present (quarks $u$ and $d$) and there are 2
quarks ungapped $q_{ub}, q_{db}$ and 4  gapped $q_{ur}$, $q_{ug}$,
$q_{dr}$, $q_{dg}$, where the color indices
 $1,2,3$ have been identified with $r,g,b$ (red, green and blue).
 The difference with the usual 2SC phase is that  color
 and electrical neutrality are required:\be \frac{\de \Omega}{\de\mu_e}=\frac{\de
\Omega}{\de\mu_3}=\frac{\de \Omega}{\de\mu_8}=0\ee This creates a
mismatch between the two Fermi spheres given by $
\delta\mu={\mu_e}/2$. Furthermore one has to satisfy the gap
equation. One finds two branches of solutions of the gap equation
corresponding to the BCS phase and to the Sarma phase. It turns out
that the solution to the present problem belongs to the Sarma
branch. In \cite{huang:2003ab} it is also shown that the solution is
a minimum of the free energy following the neutrality line. On the
other hand this point is a maximum following the appropriate line
$\mu_e=\rm{const.}$. We see that  the neutrality conditions promote
the unstable phase (Sarma) to a stable one. However this phase has
an instability connected to the Meissner mass of the gluons
\cite{huang_instability1}. In this phase the color group $SU_c(3)$
is spontaneously broken to $SU_c(2)$ with 5 of the 8 gluons
acquiring a mass; precisely the gluons 4,5,6,7,8. At the point
$\delta\mu=\Delta$ where the 2SC phase goes into the g2SC one, all
the massive gluons have imaginary mass. Furthermore the gluons
4,5,6,7 have imaginary mass already starting at
$\delta\mu=\Delta/\sqrt{2}$, that is at the Chandrasekhar-Clogston
point. This shows that both  the g2SC and the 2SC phases are
unstable. The instability of the g2SC phase seems to be a general
feature of the phases with gapless modes \cite{alford_wang}.

\section{The gCFL phase}

The gCFL phase is a generalization of the CFL phase which has been
studied both at $T=0$ \cite{Alford:2003fq,Alford:2004hz} and
$T\not=0$ \cite{gCFL_2}. The condensate has now the form \be \langle
0|\psi_{aL}^\alpha\psi_{bL}^\beta|0\rangle=\Delta_1\epsilon^{\alpha\beta
1}\epsilon_{ab1}+\Delta_2\epsilon^{\alpha\beta
2}\epsilon_{ab2}+\Delta_3\epsilon^{\alpha\beta 3}\epsilon_{ab3}\ee
The CFL phase corresponds to all the three gaps $\Delta_i$ being
equal. Varying the gaps one gets many different phases. In
particular we will be interested to the CFL, to the g2SC
characterized by $\Delta_3\not=0$ and $\Delta_1=\Delta_2=0$ and to
the gCFL phase with $\Delta_3>\Delta_2>\Delta_1$. Notice that here
in the g2SC phase the strange quark is present but unpaired.

In flavor space the gaps $\Delta_i$ correspond to the following
pairings \be \Delta_1\Rightarrow ds,~~~\Delta_2\Rightarrow us,~~~
\Delta_3\Rightarrow ud\ee The mass of the strange quark is taken
into account by shifting all the chemical potentials involving the
strange quark as follows: $\mu_{\alpha s}\to \mu_{\alpha s}
-{M_s^2}/{2\mu}$. It has also been shown in ref. \cite{alford} that
color and electric neutrality in CFL require \be
\mu_8=-\frac{M_s^2}{2\mu},~~~\mu_e=\mu_3=0\ee At the same time the
various mismatches are given by \be
\delta\mu_{bd-gs}=\frac{M_s^2}{2\mu},~~~\delta\mu_{rd-gu}=\mu_e=0,~~~
\delta\mu_{rs-bu}=\mu_e-\frac{M_s^2}{2\mu}\ee It turns out that in
the gCFL the electron density is different from zero and, as a
consequence, the mismatch between the quarks $d$ and $s$ is the
first one to give rise to the unpairing of the corresponding quarks.
This unpairing is expected to occur for \be
2\frac{M_s^2}{2\mu}>2\Delta~~
\Rightarrow~~\frac{M_s^2}{\mu}>2\Delta\ee This has been
substantiated by the calculations in a NJL model modeled on one
gluon-exchange in \cite{Alford:2004hz}. The transition from the CFL
phase, where all gaps are equal, to the gapless phase occurs roughly
at $M_s^2/\mu =2\Delta$.
\begin{figure}[h]\begin{center}
  \includegraphics[height=.25\textheight]{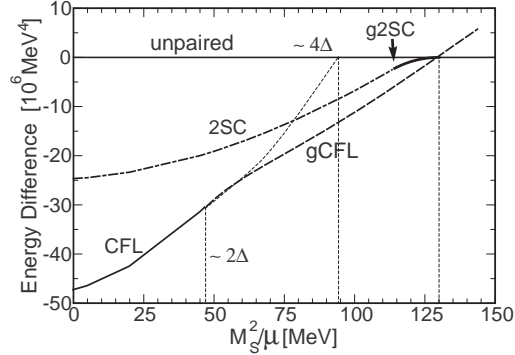}
  \caption{ We give here the free energy of the various phases with
  reference to the normal phase , named unpaired in the figure.
  \label{fig:6}}\end{center}
\end{figure}
In Fig. \ref{fig:6} we show the free energy of the various phases
with reference to the normal phase. The CFL phase is the stable one
up to $M_s^2/\mu\approx  2\Delta$. Then the gCFL phase takes over up
to about 130 $MeV$ where the system goes to the normal phase. Notice
that except in a very tiny region around this point, the CFL and
gCFL phases win over the corresponding 2SC and g2SC ones. The thin
short-dashed line represents the free energy of the CFL phase up to
the point where it becomes equal to the free-energy of the normal
phase. This happens for $M_s^2/\mu\approx 4\Delta$. This point is
the analogue  of the Chandrasekhar-Clogston point of the two-flavor
case.

Also the gCFL phase has gapless excitations and as a consequence the
chromomagnetic instability discussed in the case of the g2SC phase
shows up  here too. This has been shown in
\cite{MeissnerCFL_1,MeissnerCFL_2}.

The existence of the chromomagnetic instability is a serious problem
for the gapless phases (g2SC and gCFL) but also for the 2SC phase,
as we have discussed previously. A way out of this problem would be
to have gluon condensation. For instance, if one assumes
artificially $\langle A_\mu^3\rangle$ and $\langle A_\mu^8\rangle$
not zero and with a value of about 10 $MeV$ it can be shown that the
instability disappears \cite{MeissnerCFL_1}. Also, very recently in
\cite{miransky}, it has been shown the possibility of eliminating
the chromomagnetic instability in the 2SC phase through a gluonic
phase. However it is not clear if the same method can be extended to
the gapless phases.

Another interesting possibility has been considered in three papers
by Giannakis and Ren, who have considered the LOFF phase, that is a
nonhomogeneous phase first studied in a condensed matter context
\cite{LOFF1,LOFF2} and then in QCD in
\cite{Alford:2000ze,Bowers:2002xr} (for recent reviews of the LOFF
phase, see \cite{Casalbuoni:2003wh,Bowers:2003ye}). The results
obtained by Giannakis and Ren in the two-flavor case are the
following:
\begin{itemize}
  \item The presence of the chromomagnetic instability in g2SC is
  exactly what one needs in order  the LOFF phase to be
  energetically favored \cite{Ren1}.
  \item  The LOFF phase in the two-flavor case has no chromomagnetic
  instabilities (though it has gapless modes) at least in the weak
  coupling limit \cite{Ren2,Ren3}.
\end{itemize}
Of course these results make the LOFF phase a natural candidate for
the stable phase of QCD at moderate densities. In the next Sections
we will describe the LOFF phase in its simplest version and then a
simplified approach to the problem with three flavors will br
presented.

\section{The LOFF Phase}

According to the authors of refs. \cite{LOFF1,LOFF2} when fermions
belong to  different Fermi spheres, they  might prefer to pair
staying as much as possible close to their own Fermi surface. The
total momentum of the pair is not zero, ${\vec p}_1+{\vec p}_2=2\vec
q$ and, as we shall show, $|\vec q\,|$ is fixed variationally
whereas the direction of $\vec q$ is chosen spontaneously. Since the
total momentum of the pair is not zero the condensate breaks
rotational and translational invariance. The simplest form of the
condensate compatible with this breaking is just a simple plane wave
(more complicated possibilities will be discussed later) \be
\langle\psi(x)\psi(x)\rangle\approx\Delta\, e^{2i\vec q\cdot\vec
x}\label{single-wave}\ee It should also be noticed that the pairs
use much less of the Fermi surface than they do in the BCS case. For
instance, if both fermions are sitting at their own Fermi surface,
they can pair only if they belong to  circles fixed by $\vec q$.
More generally there is a quite large region in momentum space (the
so called blocking region) which is excluded from pairing. This
leads to a condensate generally smaller than the BCS one.

Let us now consider in more detail the LOFF phase. For two fermions
at different densities  we have an extra term in the hamiltonian
which can be written as \be
H_I=-\delta\mu\sigma_3\label{interaction}\ee where, in the original
LOFF papers \cite{LOFF1,LOFF2}, $\delta\mu$ is proportional to the
magnetic field due to the impurities, whereas in the actual case
$\delta\mu=(\mu_1-\mu_2)/2$ and $\sigma_3$ is a Pauli matrix acting
on the two fermion space. According to refs. \cite{LOFF1,LOFF2} this
favors the formation of pairs with momenta \be \vec p_1=\vec k+\vec
q,~~~\vec p_2=-\vec k+\vec q\ee We will discuss in detail the case
of a single plane wave (see eq. (\ref{single-wave})). The
interaction term of eq. (\ref{interaction}) gives rise to a shift in
$\xi$  due both to the non-zero momentum of the pair and to the
different chemical potentials \be \xi=E(\vec p)-\mu\to E(\pm\vec
k+\vec q)-\mu\mp\delta\mu\approx \xi\mp\bar\mu\ee with \be
\bar\mu=\delta\mu-{\vec v}_F\cdot\vec q\ee Notice that the previous
dispersion relations show the presence of gapless modes at momenta
depending on the angle of $\vec v_F$ with $\vec q$. Here we have
assumed $\delta\mu\ll\mu$ (with $\mu=(\mu_1+\mu_2)/2$) allowing us
to expand $E$ at the first order in $\vec q/\mu$.

The gap equation for the present case is obtained simply from eq.
(\ref{gap}) via the substitution $\delta\mu\to\bar\mu$. By studying
this equation one can  show that increasing $\delta\mu$ from zero we
have first the BCS phase. Then  at $\delta\mu=\delta\mu_1$ there is
a first order transition to the LOFF phase
\cite{LOFF1,Alford:2000ze}, and at
$\delta\mu=\delta\mu_2>\delta\mu_1$ there is a second order phase
transition to the normal phase \cite{LOFF1,Alford:2000ze}. We start
comparing the grand potential in the BCS phase to the one in the
normal phase. Their difference is given by (see for example ref.
\cite{Casalbuoni:2003wh})\be \Omega_{\rm BCS}-\Omega_{\rm
normal}=-\frac{p_F^2}{4\pi^2v_F}\left(\Delta^2_0-2\delta\mu^2\right)\ee
where the first term comes from the energy necessary to the BCS
condensation, whereas the last term arises from the grand potential
of two free fermions with different chemical potential. We recall
also that for massless fermions $p_F=\mu$ and $v_F=1$. We have again
assumed $\delta\mu\ll\mu$. This implies that there should be a first
order phase transition from the BCS to the normal phase at
$\delta\mu=\Delta_0/\sqrt{2}$ \cite{Chandrasekhar}, since the BCS
gap does not depend on $\delta\mu$. In order to compare with the
LOFF phase one can  expand the gap equation around the point
$\Delta=0$ (Ginzburg-Landau expansion)  to explore the possibility
of a second order phase transition \cite{LOFF1}. The result for the
free energy is \be \Omega_{\rm LOFF}-\Omega_{\rm normal}\approx
-0.44\,\rho(\delta\mu-\delta\mu_2)^2\ee At the same time, looking at
the minimum in $q$ of the free energy one finds \be qv_F\approx
1.2\, \delta\mu\label{q}\ee Since we are expanding in $\Delta$, in
order to get this result
 it is enough to minimize the
coefficient of $\Delta^2$ in the free-energy (the first term in the
Ginzburg-Landau expansion).

 We see that in the window
between the intersection of the BCS curve and the LOFF curve  and
$\delta\mu_2$, the LOFF phase is favored. Also at the intersection
there is a first order transition between the LOFF and the BCS
phase. Furthermore, since $\delta\mu_2$ is very close to
$\delta\mu_1$ the intersection point is practically given by
$\delta\mu_1$. The window of existence of the LOFF phase
$(\delta\mu_1,\delta\mu_2)\simeq(0.707,0.754)\Delta_0$ is rather
narrow, but there are indications that considering the realistic
case of QCD \cite{Leibovich:2001xr} the window  opens up. Such
opening occurs also for different structures than the single plane
wave\cite{Bowers:2002xr,Casalbuoni:2004wm}.

\section{The LOFF phase with three flavors}

\begin{figure}[h]\begin{center}
\includegraphics[height=.25\textheight]{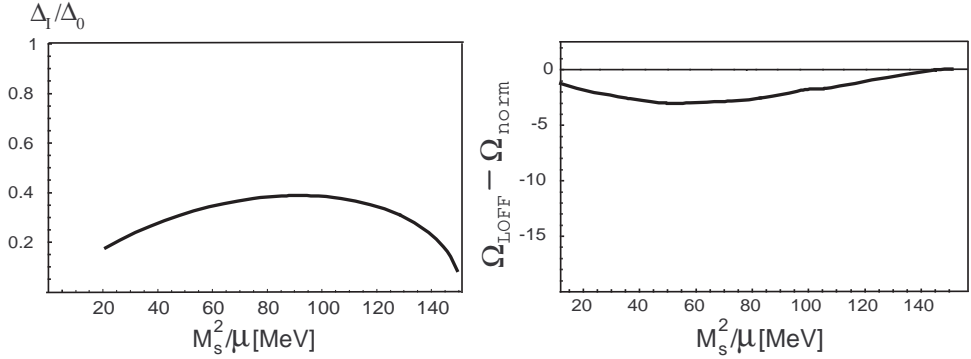}
\caption{In the left panel the gaps for LOFF with three flavors vs.
$M_s^2/\mu$. In the right panel the free energy of the most favored
solution  considered for LOFF with three flavors vs. $M_s^2/\mu$.
 \label{fig:11}}\end{center}
\end{figure}

\begin{figure}[t]\begin{center}
\includegraphics[height=.25\textheight]{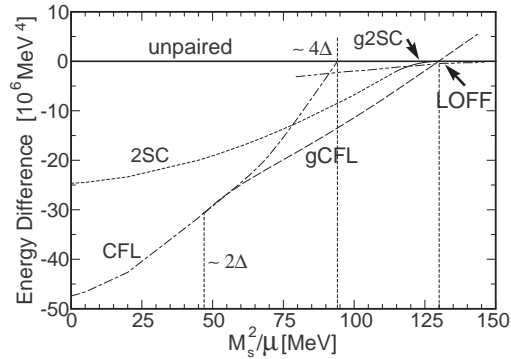}
\caption{Comparison of the free energy of the various phases already
considered in Fig. \ref{fig:6} (same notations as here) with the
LOFF phase with three flavors. \label{fig:13}}\end{center}
\end{figure}

In the last Section we would like to illustrate some preliminary
result about the LOFF phase with three flavors. This problem has
been considered in \cite{casalbuoni_loff3} under various simplifying
hypothesis:
\begin{itemize}
  \item The study has been made in the Ginzburg-Landau
  approximation.
  \item Only electrical neutrality has been required and the
  chemical potentials for the color charges $T_3$ and $T_8$ have
  been put equal to zero (see later).
  \item The mass of the strange quark has been introduced as it was
  done previously  for the gCFL phase.
  \item The study has been restricted to plane waves, assuming the
  following generalization of the gCFL case:
  \be \langle\psi^\alpha_{aL}\psi^\beta_{bL}\rangle=\sum_{I=1}^3\Delta_I(\vec x)
\epsilon^{\alpha\beta I}\epsilon_{ab I},~~~\Delta_I(\vec x)=\Delta_I
e^{2i\vec q_I\cdot\vec x}\ee
\item The condensate depends on three momenta, meaning three lengths
of the momenta $q_i$ and three angles. In \cite{casalbuoni_loff3}
only  four particular geometries have been considered: 1) all the
momenta parallel, 2) $\vec q_1$ antiparallel to $\vec q_2$ and $\vec
q_3$, 3) $\vec q_2$ antiparallel to $\vec q_1$ and $\vec q_3$, 4)
$\vec q_3$ antiparallel to $\vec q_1$ and $\vec q_2$.
\end{itemize}
The minimization of the free energy with respect to the $|\vec
q_I|$'s leads to the same result as in eq. (\ref{q}), $ |\vec
q_I|=1.2 \delta\mu_I\label{eq:55}$. Then, one has to minimize with
respect to the gaps and  $\mu_e$ in order to require electrical
neutrality. It turns out that the configurations 3 and 4 have an
extremely small gap. Furthermore for $M_s^2/\mu$ greater than about
80 $MeV$  the solution gives $\Delta_1=0$ and $\Delta_2=\Delta_3$.
In this case the configurations 1 and 2 have the same free energy.
The results for the free energy and for the gap of this solution are
given in Fig. \ref{fig:11}. In this study, the following choice of
the parameters has been made: the BCS gap, $\Delta_0=25~MeV$, and
the chemical potential $\mu=500~MeV$. The values are the same
discussed previously for gCFL in order to allow for a comparison of
the results.
We are now in the position to compare these results with the ones
obtained in \cite{Alford:2004hz} for the gCFL phase. The comparison
is made in Fig. \ref{fig:13}. Ignoring the chromomagnetic
instabilities of the gapless phases and of 2SC we see that LOFF
takes over with respect to gCFL at about $M_s^2/\mu=128~MeV$ and
goes over to the normal phase for $M_s^2/\mu\approx 150~MeV$.
However, since  the instability exists, it should be cured  in some
way. The results for the LOFF phase, assuming that also for three
flavors the chromomagnetic instability does not show up, say that
the LOFF phase could  take over the CFL phase before the transition
to gCFL. For this it is necessary that the window for the LOFF phase
gets enlarged. However, in \cite{Casalbuoni:2004wm} it has been show
that for structures more general than the plane wave the windows may
indeed becomes larger. If we define the window for the single plane
wave as $(\delta\mu_2-\delta\mu_1)/\delta\mu_2$ (see the previous
Section) we would get 0.06. The analogous ratio in going from one to
three plane waves goes to about $(150-115)/150=.23$, with a gain of
almost a factor 4. On the other hand, in \cite{Casalbuoni:2004wm} it
has been shown that considering some of the crystalline structures
already taken in exam in \cite{Alford:2000ze}, as the face centered
cube or the cube the window becomes  $(1.32-0.707)/1.32=0.46$ with a
gain of about 7.7 with respect to the single plane wave. If these
gains would be maintained in going from two to three flavors with
the face centered cube structure, one could expect a gain from 4 to
7.7 with an enlargement of the window between 88 and 170 $MeV$,
which would be enough to cover the region of gCFL (which is about 70
$MeV$). At last we want to comment about the approximation in
neglecting the color neutrality condition and assuming
$\mu_3=\mu_8=0$. The results of ref. \cite{casalbuoni_loff3} show
that $\mu_e\approx M_s^2/(4\mu)$ as for the case of 3 color and 3
flavor unpaired quarks \cite{alford}. Furthermore the unpaired
quarks have also $\mu_3=\mu_8=0$. Also, from Fig. \ref{fig:11} we
see that in our approximations the transition from the LOFF to the
normal phase is very close to be continuous. Since we expect  the
chemical potentials to be continuous at the transition point, we
expect $\mu_3=\mu_8=0$ also on the LOFF side, at least when close to
the critical point. This means the color neutrality condition should
be $\mu_3=\mu_8=0$ in the neighborhood of the transition. Therefore
we expect  the determination of the point $M_s^2/\mu=150~MeV$ to be
safe. On the other hand, the requirement of color neutrality could
change the intersection point with gCFL. Nevertheless, since the
critical point for LOFF is higher than the one of gCFL, for
increasing $M_s$ the system must to go into the LOFF phase.



\begin{thebibliography}{99}



\bibitem{barrois}

B. Barrois, {\it Nuclear Physics} {\bf B129}, 390 (1977);
S. Frautschi, {\it Proceedings of workshop on hadronic matter at
extreme density}, Erice 1978; D. Bailin and A. Love, {\it Physics
Report} {\bf 107} (1984) 325 .


\bibitem{cs} M.~Alford, K.~Rajagopal, and F.~Wilczek,
{\it Phys.\ Lett.}\  {\bf B422}(1998) 247 [hep-ph/9711395];
R.~Rapp, T.~Schafer, E.~V.~Shuryak and M.~Velkovsky,
\textit{Phys.\ Rev.\ Lett.}\  {\bf 81}, 53 (1998) [hep-ph/9711396].


\bibitem{weak}  D.T.~Son,
{\it Phys.\ Rev.}\  {\bf D59} (1999) 094019 [{hep-ph/9812287}];
T.~Sch\"{a}fer and F.~Wilczek, {\it Phys.\ Rev.}\  {\bf D60} (1999)
114033 [{hep-ph/9906512}];
D.K.~Hong, V.A.~Miransky, I.A.~Shovkovy, and L.C.R.~Wijewardhana,
{\it Phys.\ Rev.}\  {\bf D61} (2000) 056001 [{hep-ph/9906478}];
S.D.H.~Hsu and M.~Schwetz, {\it Nucl.\ Phys.}\ {\bf B572} (2000) 211
[{hep-ph/9908310}];
W.E.~Brown, J.T.~Liu, and H.-C.~Ren, {\it Phys.\ Rev.}\  {\bf D61}
(2000) 114012 [{hep-ph/9908248}].



\bibitem{PR-sp1} R.D.~Pisarski and D.H.~Rischke,
{\it Phys.\ Rev.}\  {\bf D61} (2000) 051501 [{nucl-th/9907041}].

\bibitem{weak-cfl} I.A.~Shovkovy and L.C.R.~Wijewardhana,
{\it Phys.\ Lett.}\  {\bf B470} (1999) 189 [{hep-ph/9910225}];
T.~Sch\"{a}fer, {\it Nucl.\ Phys.}\ {\bf B575} (2000) 269
[{hep-ph/9909574}].

\bibitem{Amore:2001uf}
P.~Amore, M.~C.~Birse, J.~A.~McGovern and N.~R.~Walet,
{\it Phys.\ Rev.}\  {\bf D65} (2002) 074005 [{hep-ph/0110267}].

\bibitem{Casalbuoni:2003wh}
  R.~Casalbuoni and G.~Nardulli,
  {\it Rev.\ Mod.\ Phys.}\  {\bf 76} (2004) 263
  [{hep-ph/0305069}].

\bibitem{sarma:1963sa}
 G.~Sarma, {\it J. Phys. Chem. Solids} {\bf 24} (1963) 1029.

\bibitem{Chandrasekhar}
B.~S.~Chandrasekhar, {\it App. Phys. Lett.} {\bf 1} (1962) 7;
A.~M.~Clogston, {\it Phys. Rev. Lett.} {\bf 9} (1962) 266.

\bibitem{huang:2003ab}
I.~Shovkovy and M.~Huang,
{\it Phys.\ Lett.}\  {\bf B564} (2003) 205 [{hep-ph/0302142}].


\bibitem{huang_instability1}
M. Huang and I. A. Shovkovy, {\it Phys. Rev.} {\bf D70} (2004)
051501 [{hep-ph/0407049}]; {\it ibidem} {\it Phys. Rev.} {\bf D70}
(2004) 094030 [{hep-ph/0408268}].




\bibitem{alford_wang}
M. Alford and Q. Wang, {\it J. Phys. } {\bf G31} (2005) 719
[{hep-ph/0501078}].


\bibitem{Alford:2003fq}
M.~Alford, C.~Kouvaris and K.~Rajagopal,
{\it Phys.\ Rev.\ Lett.}\  {\bf 92} (2004) 222001
[{hep-ph/0311286}].

\bibitem{Alford:2004hz}
M.~Alford, C.~Kouvaris and K.~Rajagopal,
{\it Phys.\ Rev.} {\bf D71} (2005) 054009 [{hep-ph/0406137}].

\bibitem{gCFL_2}
M. Alford, P. Jotwani, C. Kouvaris, J. Kundu and K. Rajagopal, {\it
Phys. Rev.} {\bf D71} (2005) 114011 [{astro-ph/0411560}].

\bibitem{alford}
M. Alford and K. Rajagopal, {\it JHEP} {\bf 06} (2002) 031
[{hep-ph/0204001}].


\bibitem{MeissnerCFL_1}
R. Casalbuoni,  R. Gatto,  M. Mannarelli, G. Nardulli and M.
      Ruggieri,  {\it Phys. Lett.}
      {\bf B605} (2005) 362 [{hep-ph/0410401}].

\bibitem{MeissnerCFL_2}
K. Fukushima, {\it Phys. Rev.} {\bf D70} (2005) 07002
[{hep-ph/0506080}].


\bibitem{miransky}
E.V. Gorbar, Michio Hashimoto and V.A. Miransky,  {hep-ph/0509334}.


\bibitem{LOFF1}
A.~I.~Larkin and Yu.~N.~Ovchinnikov,  {\it Sov. Phys. JETP} {\bf 20}
(1965) 762.

\bibitem{LOFF2} P.~Fulde and R.~A.~Ferrell, {\it Phys. Rev.} {\bf 135}  (1964) A550.



\bibitem{Alford:2000ze} M.~G.~Alford,
J.~A.~Bowers and K.~Rajagopal,
{\it Phys.\ Rev.}  {\bf D63}  (2001) 074016 [{hep-ph/0008208}].

\bibitem{Bowers:2002xr}
J.~A.~Bowers and K.~Rajagopal,
{\it Phys.\ Rev.} {\bf D66}  (2002) 065002 [{hep-ph/0204079}].


\bibitem{Bowers:2003ye}
J.~A.~Bowers, {hep-ph/0305301}.

\bibitem{Ren1}
I. Giannakis and H.C. Ren, {\it Phys. Lett.} {\bf B611} (2005) 137
[{hep-ph/0412015}].

\bibitem{Ren2}
I. Giannakis and H.C. Ren, {\it Nucl. Phys.} {\bf B723} (2005) 255
[{hep-th/0504053}].

\bibitem{Ren3}
I. Giannakis and H.C. Ren, {\it Phys. Lett.} {\bf B631} (2005) 16
[{hep-ph/0507306}].


\bibitem{Leibovich:2001xr}
A.~K.~Leibovich, K.~Rajagopal and E.~Shuster,
{\it Phys.\ Rev.} {\bf D64}  (2001) 094005  [{hep-ph/0104073}]; see
also I.~Giannakis, J.~T.~Liu and H.~C.~Ren,
{\it Phys.\ Rev.} {\bf D66}  (2002) 031501  [{hep-ph/0202138}].




\bibitem{casalbuoni_loff3}
R. Casalbuoni,  R. Gatto, N. Ippolito, G. Nardulli, and M. Ruggieri,
{\it Phys. Lett.} {\bf B627} (2005) 89 [{hep-ph/0507247}]; see also
Erratum, to be published.

\bibitem{Casalbuoni:2004wm}
R.~Casalbuoni, M.~Ciminale, M.~Mannarelli, G.~Nardulli, M.~Ruggieri
and R.~Gatto, {\it Phys. Rev.} {\bf D70} (2004) 054004
[{hep-ph/0404090}].

\end{thebibliography}
\end{document}